\documentclass[12pt]{article}
\usepackage{amssymb,amsmath,mathrsfs}
\usepackage{hyperref}
\usepackage{graphicx}
\usepackage{color}
\usepackage[OT2,OT1]{fontenc}
\usepackage{upquote}
\usepackage{authblk}
\usepackage[numbers,sort&compress]{natbib}
\usepackage{setspace}
\usepackage{float}

\begin{document}

\title{\vspace{-1cm}\Large\bf Effect of the instrument slit function on upwelling radiance from a wavelength dependent surface reflectance}

\author[1, 2]{\small Rajinder K. Jagpal}
\author[1, 2, 3]{\small Rehan Siddiqui}
\author[1, 3, 4]{\small Sanjar M. Abrarov}
\author[2, 3, 4]{\small \\ Brendan M. Quine}

\affil[1]{\scriptsize Epic College of Technology, 5670 McAdam Rd., Mississauga, Canada, L4Z 1T2 \normalsize}
\affil[2]{\scriptsize Dept. Physics and Astronomy, York University, 4700 Keele St., Toronto, Canada, M3J 1P3 \normalsize}
\affil[3]{\scriptsize Dept. Earth and Space Science and Engineering, York University, 4700 Keele St., Canada, M3J 1P3 \normalsize}
\affil[4]{\scriptsize Thoth Technology Inc., Algonquin Radio Observatory, Achray Road, RR6, Pembroke, Ontario, Canada, K8A 6W7 \normalsize}

\date{October 11, 2022}
\maketitle

\vspace{-0.75cm}

\begin{abstract}
The Radiance Enhancement (RE) method was introduced for efficient detection of clouds from the space. Recently, we have also reported that due to high reflectance of combustion-originated smokes, this approach can also be generalized for detection of the forest fires by retrieving and analyzing datasets collected from a space orbiting micro-spectrometer operating in the near infrared spectral range. In our previous publication, we have performed a comparison of observed and synthetic radiance spectra by developing a method for computation of surface reflectance consisting of different canopies by weighted sum based on their areal coverage. However, this approach should be justified by a method based on 
corresponding proportions of the upwelling radiance. The results of computations we performed in this study reveal a good match between areal coverage of canopies and the corresponding proportions of the upwelling radiance due to effect of the instrument slit function.
\vspace{0.25cm}
\\
\noindent {{\bf Keywords}: Radiance Enhancement, upwelling radiance, line-by-line computation, radiative transfer model}  \\
\end{abstract}

\section{Introduction}

The detection of the upwelling radiance (also known as the flux) from space, especially in the near infrared (NIR) spectral regions, plays an important role in Atmospheric Science. In particular, this technique provides data that can be used to retrieve valuable information about atmospheric constituents of the gases and condition of the surface. The large-scale datasets collected from space by a micro-spectrometer orbiting around the Earth can be retrieved to obtain the mixing ratios of the atmospheric gases like $\rm{CO_2}$, $\rm{CO}$, $\rm{H_{2}O}$, $\rm{CH_4}$, $\rm{O_2}$, $\rm{O_3}$, $\rm{N_{2}O}$ \cite{Buchwitz2000, Buchwitz2005a, Buchwitz2005b, Bosch2006, Jagpal2010, Jagpal2011} and to estimate levels of particulate matter, especially hazardous $\rm{PM_{2.5}}$ particles \cite{Christopher2010}. This type of large-scale remote space data are used by our research group to detect and monitor the sources of some greenhouse gases \cite{Jagpal2010, Jagpal2011, Jagpal2019, Siddiqui2015, Siddiqui2017a}. The retrieved results of the NIR space data can be used not only to analyze and trace greenhouse gases, but also to predict their tendency and dynamics.

We have shown recently that the datasets of the upwelling radiance collected from space remote sensor can also be used for efficient detection of cloud scenes by using the Radiance Enhancement method \cite{Siddiqui2017b, Siddiqui2020a, Siddiqui2020b, Siddiqui2021}. Specifically, the RE method utilizes the line-by-line radiative transfer model GENSPECT \cite{Quine2002} in order to match the synthetic radiance with observed radiance by incrementing or decrementing several variables like concentration of gases, zenith angle, deviation of the nadir angle, etc. The computation is performed in a nested loop until a best match of synthetic and observed radiance spectra is achieved. A least square method is applied as a criterion to match the computed and observed data \cite{Siddiqui2021}. Apart from the variable parameters the wavelength dependency of the reflecting surface has to be taken into consideration.

The RE method is based on the following two formulas \cite{Siddiqui2017b, Siddiqui2020a, Siddiqui2020b, Siddiqui2021}
$$
R{E_i} = \frac{1}{N}\sum\limits_{j = 1}^J {\left\{ {\frac{{OB{S_i}\left[ j \right] - SY{N_i}\left[ j \right]}}{{SY{N_i}\left[ j \right]}}} \right\},}
$$
and
$$
CRE = \sum\limits_{i = 1}^N {R{E_i}},
$$
where $i$ is the index of wavelength sub-bands, $j$ is the index of grid-points, $J$ is the total number of the grid-points and $N$ is the number of sub-bands that can be taken as $4$. The first formula signifies RE associated with sub-band $i$ while the second formula defines the combined radiance enhancement (CRE) for all considered sub-bands.

The RE method utilizes datasets obtained from the ultra-light and small-size space-orbiting Argus 1000 micro-spectrometer that was launched into space from India in 2008 \cite{Jagpal2011} as a payload of the CanX-2 nano-satellite \cite{Rankin2005}. This space instrument operates in the NIR range from $1100\, nm$ to $1700\, nm$ with spatial resolution on the ground about $1400 \times 1100\;{m^2}$ \cite{Jagpal2011}. By computing frequency of events as a function of the CRE, one can estimate the chances for observation of the cloud scenes. The detailed description of the RE method is described in the work \cite{Siddiqui2017b}.

Since the RE method utilizes the Argus datasets, it is developed to cover the wide NIR range from $1100\, nm$ to $1700\, nm$. Therefore, due to wide spectral range it takes into the consideration the wavelength dependency of the surface reflectance. However, if the Earth surface is not homogeneous, then the cumulative effect of upwelling radiance from each surface component must be considered. The landscapes over Canada include prairies, forests, lakes, rivers and mixed lands. The prairies are predominantly covered with grass and soil while the forests are covered with pine trees, broadleaf trees and bushes. There are vast mixed lands in North America that may include each component in different proportions \cite{Toot2020, Anderson2006}.

In order to resolve the problem of surface Albedo inhomogeneity, we applied a method based on areal coverage of each specific canopy \cite{Siddiqui2021}. However, this method has not been correlated with corresponding proportions of the upwelling radiance contributed by each type of canopy from the ground. Therefore, the computations we performed in the retrieval in our recent work \cite{Siddiqui2021} should be rigorously justified by this correlation. In this work we show the numerical evidence that fully justifies approach based on areal coverage \cite{Siddiqui2021}. In particular, the numerical analysis we performed reveals that the instrument slit function plays a key role in such a correlation between methods based on areal coverage and corresponding proportions of the upwelling radiance.

\section{Results and discussion}

\subsection{Motivation}

Since field of view of the Argus 1000 instrument covers the area about $1.56\,k{m^2}$ \cite{Jagpal2011}, it is very likely to expect that the upwelling radiance originates from inhomogeneous surface Albedo. In general, therefore, the upwelling radiance is supposed to be due to combined radiances from different canopies. Considering nadir view \cite{Jagpal2011}, we may expect that the soil, pine trees, vegetation (broadleaf trees and bushes) and grass may be the major contributors for the reflectance.
\begin{figure}[H]
\begin{center}
\includegraphics[width=20pc]{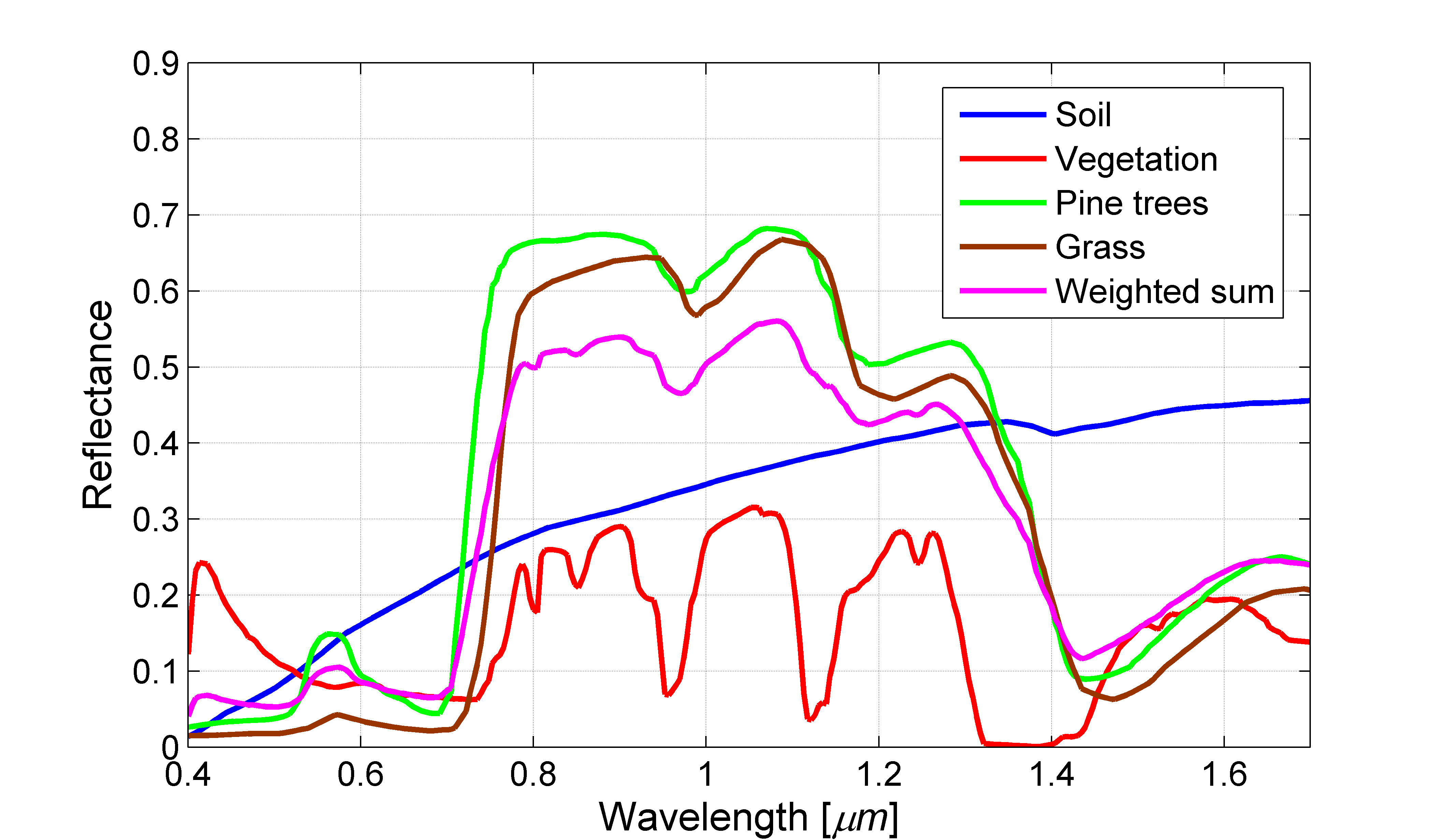}\hspace{2pc}%
\begin{minipage}[b]{26pc}
\vspace{0.3cm}
{\sffamily {\bf{Fig. 1.}} Wavelength dependent reflectance for different surfaces \cite{Roberts2013, Li2018, MODIS, Baldridge2009}.}
\end{minipage}
\end{center}
\end{figure}

Figure 1 shows reflectance as a function of the wavelength for the soil, pine trees, vegetation and grass. The data for reflectance spectra can be gathered from the following spectral data sources \cite{Roberts2013, Li2018, MODIS, Baldridge2009}. We can consider, for example, the areal coverage beneath field of view of Argus instrument for these canopies to be $13.33\% $, $20\% $, $40\% $ and $26.67\% $,  respectively. This is a typical observation for Subarctic zones of Canada.

The chill weather and clean environment are very vital factors for successful reproduction and growth of the pine trees \cite{Dick2014}. However, due to increasing atmospheric temperature appearing as a result of rapid increase of the $\rm{CO_2}$ greenhouse gas over last decades \cite{Apadula2019, Karnauskas2020}, increased level of industrial pollutants \cite{Davidson2020} and wildfires \cite{Tymstra2020, Axelson2009, Benali2016}, the broadleaf plants competing with pine trees have more chances to replace them with progressive rate. Pine trees provide the most valuable wood in construction industry. Therefore, these trees are logged in a commercially colossal scales for internal and worldwide trade. Pine trees are very slow in growth and, unlike many other species, they need several decades for recovery. As a result, we can see a significant domination of the broadleaf plants especially near the farms and villages in the rural areas of Canada. As a consequence, the areal coverage proportion of broadleaf trees and bushes to pine trees becomes larger year by year and in many places like the Wood Buffalo National Park of Canada and Algonquin National Park of Canada. Unfortunately, such a tendency becomes normal even at Northern Subarctic of Canada, where cold weather and clean environmental conditions were considered ideal for growth of the pine trees in the recent past \cite{Dick2014}.

As the increase of broadleaf plants appearing as a result of increasing atmospheric temperature is a current issue, we may attempt to analyze how the reflectance may change as a result of the increasing contribution of the vegetation. Considering that the relative areal coverage proportions of the soil, pine trees and grass remains same and increasing the contribution of the vegetation from $0$ to $1$, we can compute the reflectance as a function of the contribution factor of vegetation.
\begin{figure}[H]
\begin{center}
\includegraphics[width=20pc]{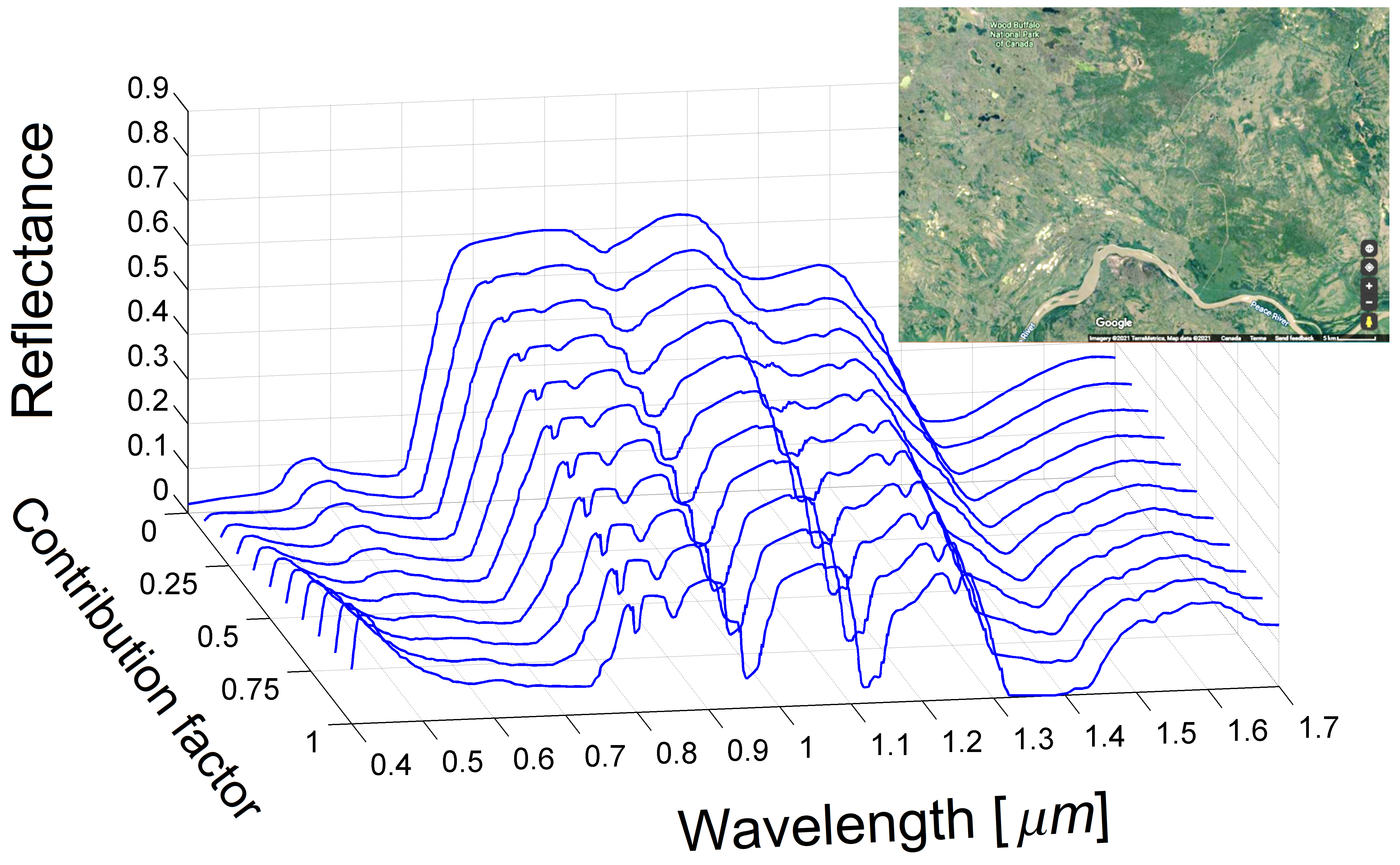}\hspace{2pc}%
\begin{minipage}[b]{26pc}
\vspace{0.3cm}
{\sffamily {\bf{Fig. 2.}} Evolution of the reflectance as a function of the contribution factor. Inset shows a typical landscape area over the Wood Buffalo National Park of Canada obtained from Google Map \cite{GE}.}
\end{minipage}
\end{center}
\end{figure}

The results of computation in illustrated in the Fig. 2 showing how the increase of vegetation variates the cumulative reflectance. Inset in Fig. 2 shows the Wood Buffalo National Park of Canada, obtained from Google Map \cite{GE}. As we can see, the surface Albedo is not uniform and mostly consists of soil, pine trees, vegetation and grass. By taking the areal coverage proportions as $13.33\% $, $20\% $, $40\% $ and $26.67\% $, respectively, we can compute the cumulative radiance by weighted sum. This signifies that parts of instrument's field of view $1.56\,km^2$ occupied by soil, pine trees, vegetation and grass are $0.1333$, $0.2$, $0.4$ and $0.2667$, respectively. The curve of cumulative reflectance obtained by weighted sum can be seen in Fig. 1, shown by magenta color.
In order to simulate the upwelling radiance we applied an updated version of the line-by-line (LBL) radiative transfer model GENSPECT \cite{Quine2002} with the HITRAN molecular spectroscopic database \cite{Hill2016}. In the updated version we applied a single domain interpolation of the Voigt function \cite{Abrarov2019} in which any of our three algorithms \cite{Abrarov2011, Abrarov2018a} or \cite{Abrarov2018b} can be implemented to generate highly accurate references. This new method accelerates the computation and, in contrast to traditional method of LBL computation, enables us to avoid unnecessary interpolation of the absorption coefficients \cite{Fomin1995, Sparks1997}. Furthermore, we developed some additional MATLAB library files in order to perform computation of the cumulative weighted sum within a wide spectral range.
\begin{figure}[H]
\begin{center}
\includegraphics[width=20pc]{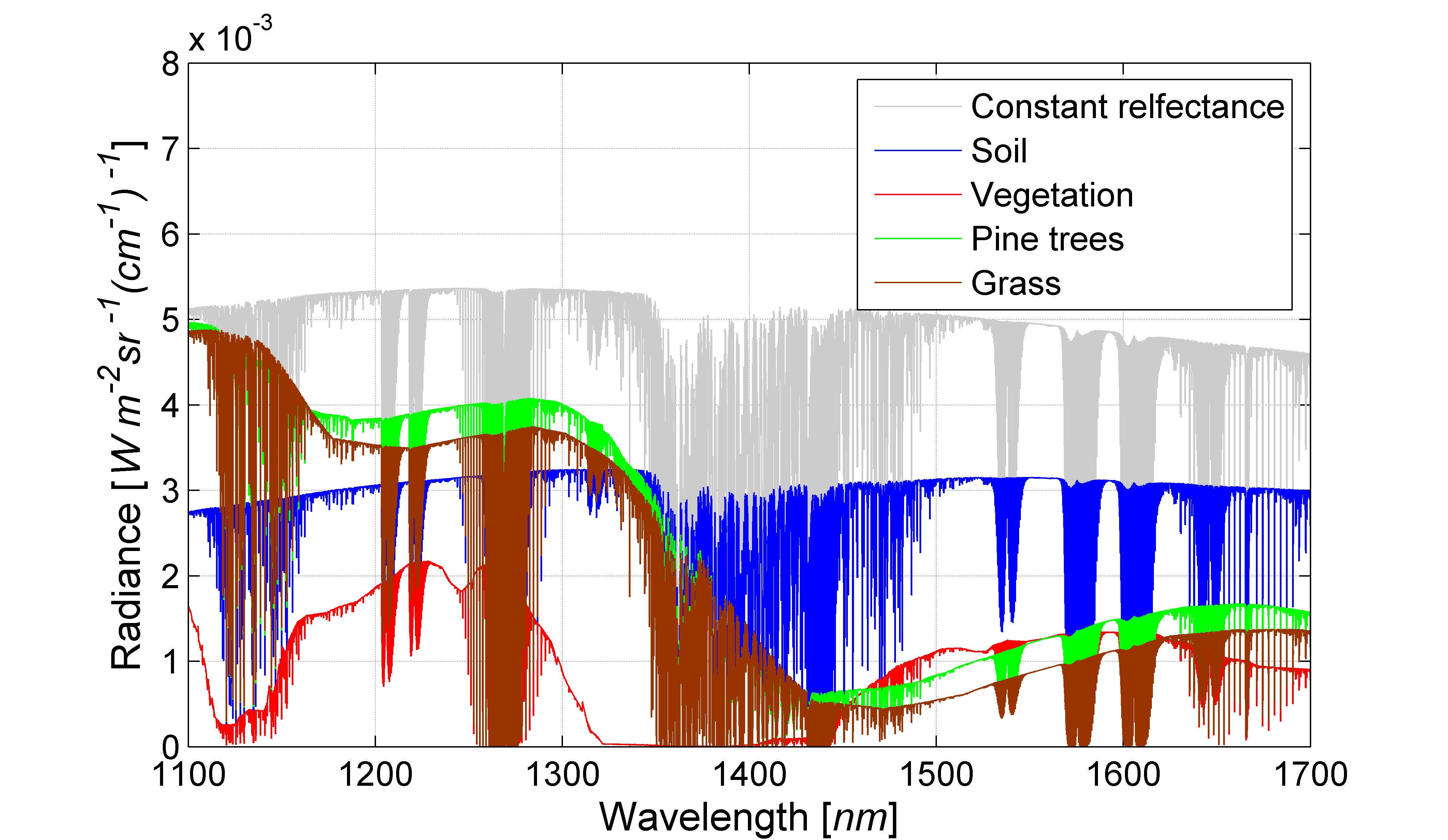}\hspace{2pc}%
\begin{minipage}[b]{26pc}
\vspace{0.3cm}
{\sffamily {\bf{Fig. 3.}} Synthetic radiance spectra upwelling from the different surfaces.}
\end{minipage}
\end{center}
\end{figure}

In the simplest case we can take surface reflectance to be a constant. This is shown in the Fig. 3 by gray highly oscillated curve. However, such a simplification does not describe the real scenario when we deal with a very wide spectral range. In case of the wavelength dependency of the surface Albedo, the shape of the relative positions of the spectral lines can significantly change. This effect of the wavelength dependency of reflectance can be seen from Fig. 3 for soil, pine trees, vegetation and grass by blue, red, green and brown highly oscillated curves, respectively.

Due to a limited resolution of any spectrometer the high-frequency oscillation in upwelling radiance 
cannot be observed. In order to account for this limitation of the instrument we took into consideration 
the instrument slit function. The effect of the instrument slit function \cite{Beirle2017, Galan1968, Roseler1966} can be seen from the Fig. 4 showing the upwelling radiance for soil, pine trees, vegetation and grass by blue, red, green and brown curves, respectively. The gray curve in this figure shows the upwelling radiance at constant reflectance. It should be noted that the shapes of these curves somehow resemble the shapes of the reflectance curves shown in Fig. 1. Therefore, we may suggest that the radiance data collected from space by micro-spectrometers like Argus 1000 can also be used to retrieve wavelength dependent reflectance spectra by calibration.

\begin{figure}[H]
\begin{center}
\includegraphics[width=20pc]{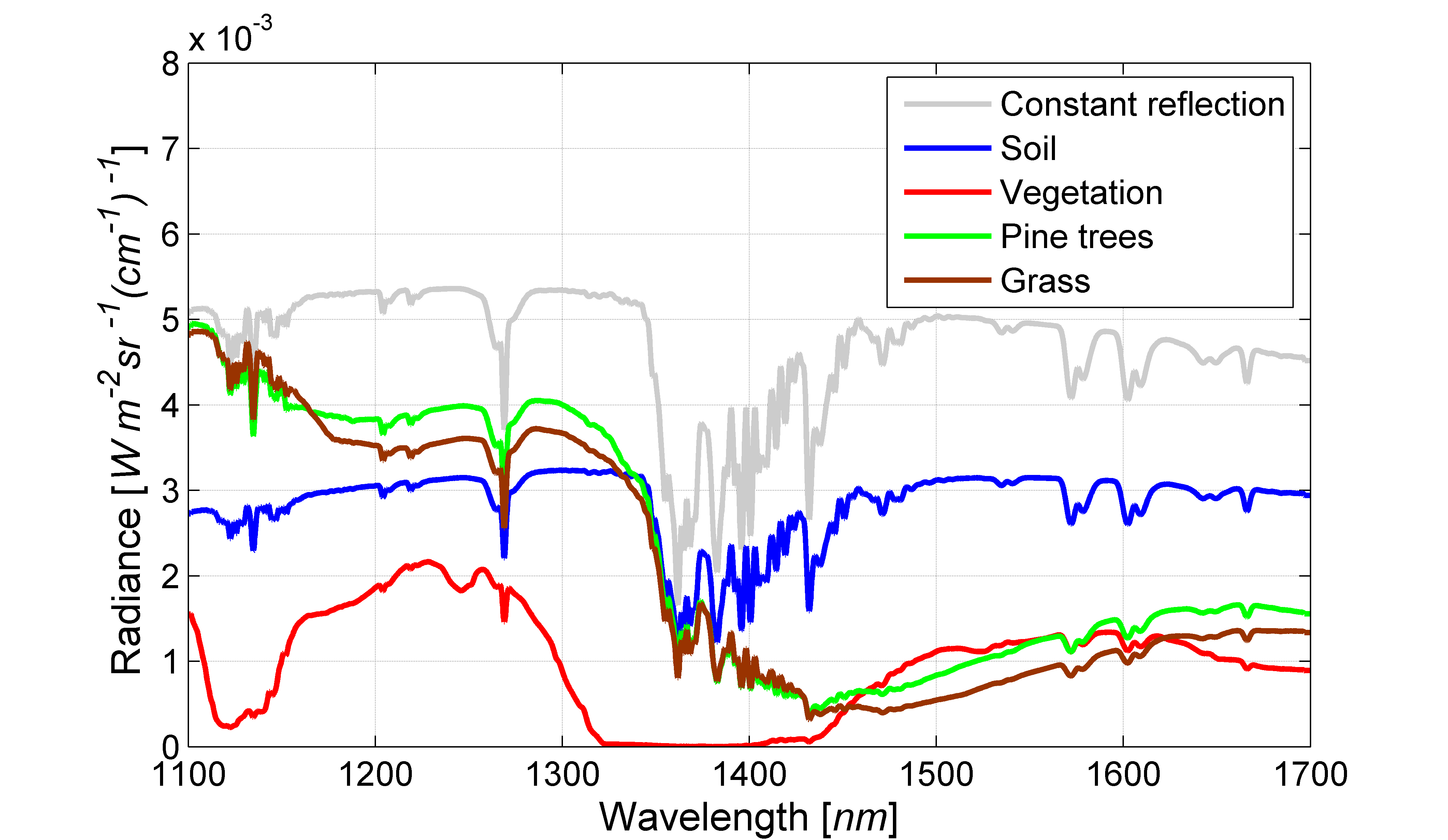}\hspace{2pc}%
\begin{minipage}[b]{26pc}
\vspace{0.3cm}
{\sffamily {\bf{Fig. 4.}} Synthetic radiance with instrument slit function for the different surfaces.}
\end{minipage}
\end{center}
\end{figure}

\begin{figure}[ht]
\begin{center}
\includegraphics[width=20pc]{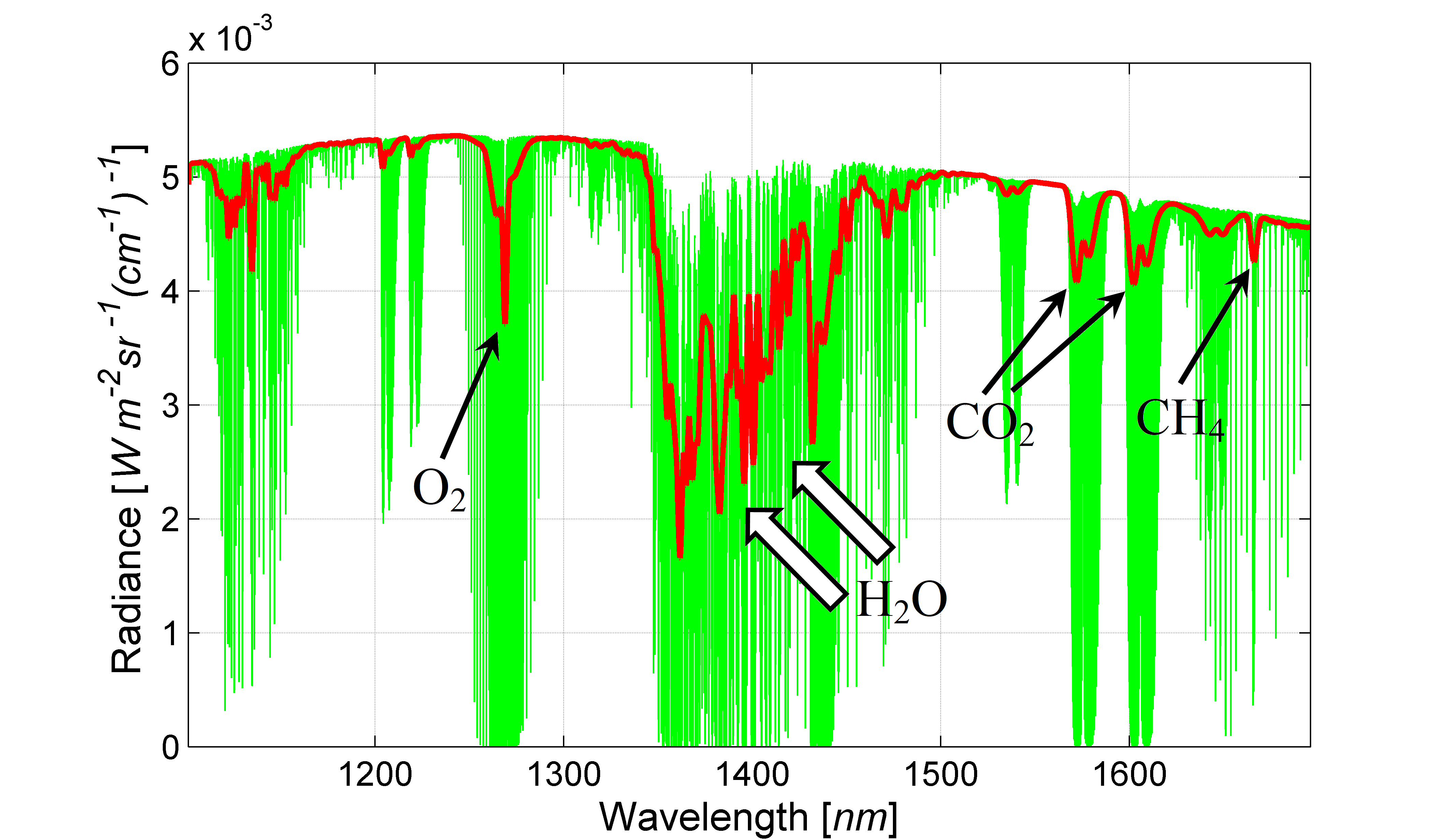}\hspace{2pc}%
\begin{minipage}[b]{26pc}
\vspace{0.3cm}
{\sffamily {\bf{Fig. 5.}} Synthetic radiance upwelling from a surface with wavelength independent reflectance.}
\end{minipage}
\end{center}
\end{figure}

The Argus 1000 micro-spectrometer is designed to observe primarily $\rm{CO_2}$ greenhouse gas. However, other three greenhouse gases $\rm{H_{2}O}$, $\rm{O_2}$ and $\rm{CH_4}$ are also of the great interest. Such a choice of these particular four gases is due to their profound absorptions in the Argus remote sensor wavelength range $1100\,nm$ to $1700\,nm$ \cite{Jagpal2011}. These absorption lines are most profound when the reflectance is a constant as shown in the Fig. 5. In particular, we can observe several sub-bands; relatively narrow sub-band near $1275\,{nm}$ is due to $\rm{O_{2}}$ molecules, wide sub-band starting from approximately $1350\,{nm}$ to $1500\,{nm}$ is due to water vapor $\rm{H_{2}O}$, two close sub-bands near $1575\,{nm}$ and $1600\,{nm}$ are due to $\rm{CO_{2}}$ and a narrow sub-band at $1675\,{nm}$ is due to $\rm{CH_4}$.

The radiance $R\left( {\nu ,{z_{obs}}} \right)$, where $\nu$ is spatial frequency and $z_{obs}$ is the observation point from space, can be defined as power per square meter per steradian per (spatial) frequency ([$Wm^{-2}sr^{-1}(cm^{-1})^{-1}$]). The radiance upwelling from the Earth surface accounts for both, reflected and emitted radiance. The upwelling radiance from the surface canopies may occur as a result of photon absorption/re-emission and reflectance. In general, the computation of the radiance may be complicated due to inhomogeneity of atmospheric column. The inhomogeneity includes multiple scattering, pressure and temperature variations with height, mixing ratio of the gases and so on. One of the efficient ways to resolve this problem is to slice the atmospheric column into cells such that each cell can be regarded as pseudo-homogeneous. Although slicing atmospheric column into layers (or cells) makes the computation more intense, it resolves many complexities and provides more realistic and reliable results. The technical aspects in computation of the radiance by a LBL radiative transfer atmospheric model can be found elsewhere \cite{Edwards1992, Liou2002, Edwards1987, Edwards1988, Nordebo2021}.

Our preference for the unit $\left[Wm^{-2}sr^{-1}(cm^{-1})^{-1}\right]$ involving spatial frequency is due to classic work of Edwards \cite{Edwards1992}, where an efficient LBL application for a radiative transfer model was developed and implemented. Furthermore, the similar units in energy-related parameters, expressed in terms of spatial frequency $cm^{-1}$, remains common in many spectroscopic applications \cite{Nordebo2021}.

In our recent publication we performed computation of the upwelling radiance by using a method based on areal coverage \cite{Siddiqui2021}. However, strictly saying this method is rather heuristic and, therefore, requires computational error analysis to validate it. Specifically, this method should be matched with that of based on a corresponding upwelling radiance proportions.

\subsection{Methodology of computation}

A method based on areal coverage takes area proportions of the soil, pine trees, vegetation and grass within field of view of the space instrument. Based on these results the cumulative wavelength dependent reflectance ${r_c}\left( \lambda  \right)$ is computed as a weighted sum that for our specific case is given by
$$
{r_c}\left( \lambda  \right) = {w_1}{r_s}\left( \lambda  \right) + w{_2}{r_{pt}}\left( \lambda  \right) + {w_3}{r_v}\left( \lambda  \right) + {w_4}{r_g}\left( \lambda  \right),
$$
where ${w_1} = 0.1333$, ${w_2} = 0.2$, ${w_3} = 0.4$ and ${w_4} = 0.2667$ are weighted coefficients such that ${w_1} + w{_2} + {w_3} + {w_4} = 1$ while ${r_s}\left( \lambda  \right)$, ${r_{pt}}\left( \lambda  \right)$, ${r_v}\left( \lambda  \right)$ and ${r_g}\left( \lambda  \right)$ are reflectance of the soil, pine trees, vegetation and grass dependent on wavelength $\lambda $. Once the cumulative reflectance is found, the LBL computation can be performed for upwelling radiance. As we can see from Fig. 1 reflectance functions from pine tree and grass resemble to each other. Therefore, we should not expect that their proportions would variate significantly the upwelling radiance. Also the reflectance curve of the soil is relatively flat within the range from $1100\,nm$ to $1700\,nm$. That signifies that its contribution would also not affect much the shape of the upwelling radiance. However, the reflectance of the vegetation is significantly different; it is not flat and does not resemble any other curves. Consequently, the contribution factor from the vegetation may have the largest impact to the upwelling radiance.

A method based on upwelling radiance proportions is different. Instead of computing cumulative reflectance by using the areal coverage, it utilizes following weighted sum formula
$$
R\left( {\nu ,{z_{obs}}} \right) = {w_1}{R_s}\left( {\nu ,{z_{obs}}} \right) + {w_2}{R_{pt}}\left( {\nu ,{z_{obs}}} \right) + {w_3}{R_v}\left( {\nu ,{z_{obs}}} \right) + {w_4}{R_g}\left( {\nu ,{z_{obs}}} \right),
$$
where ${R_s}\left( {\nu ,{z_{obs}}} \right)$, ${R_{pt}}\left( {\nu ,{z_{obs}}} \right)$, ${R_v}\left( {\nu ,{z_{obs}}} \right)$ and ${R_g}\left( {\nu ,{z_{obs}}} \right)$ are radiance functions due to contribution of soil, pine trees, vegetation and grass, respectively.

The first method of computation is more practical as it is easier to implement and it takes less amount of time for computation. The second method, however, is more rigorous as it accounts for the contribution of each component of the ground Albedo. Ideally, when ${r_s}\left( \lambda  \right)$, ${r_{pt}}\left( \lambda  \right)$, ${r_v}\left( \lambda  \right)$ and ${r_g}\left( \lambda  \right)$ are wavelength independent, then both methods provide the same result. However, because of wavelength dependence of the surface Albedo, one should expect some discrepancies in computation of the upwelling radiance and evaluate numerically whether or not these discrepancies are negligible.
\begin{figure}[H]
\begin{center}
\includegraphics[width=20pc]{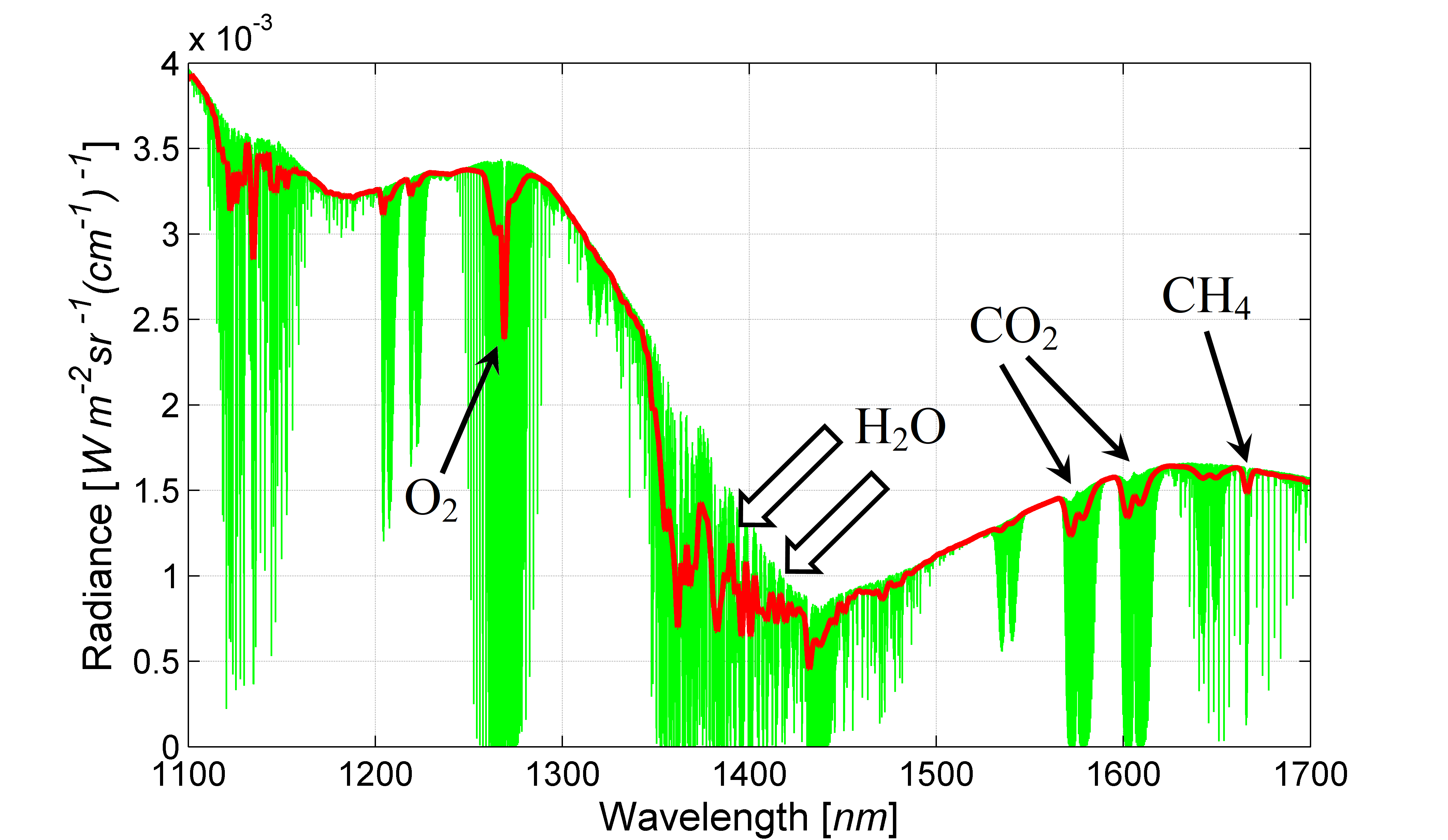}\hspace{2pc}%
\begin{minipage}[b]{26pc}
\vspace{0.3cm}
{\sffamily {\bf{Fig. 6.}} Radiance computed by using the areal coverage method.}
\end{minipage}
\end{center}
\end{figure}

\begin{figure}[H]
\begin{center}
\includegraphics[width=20pc]{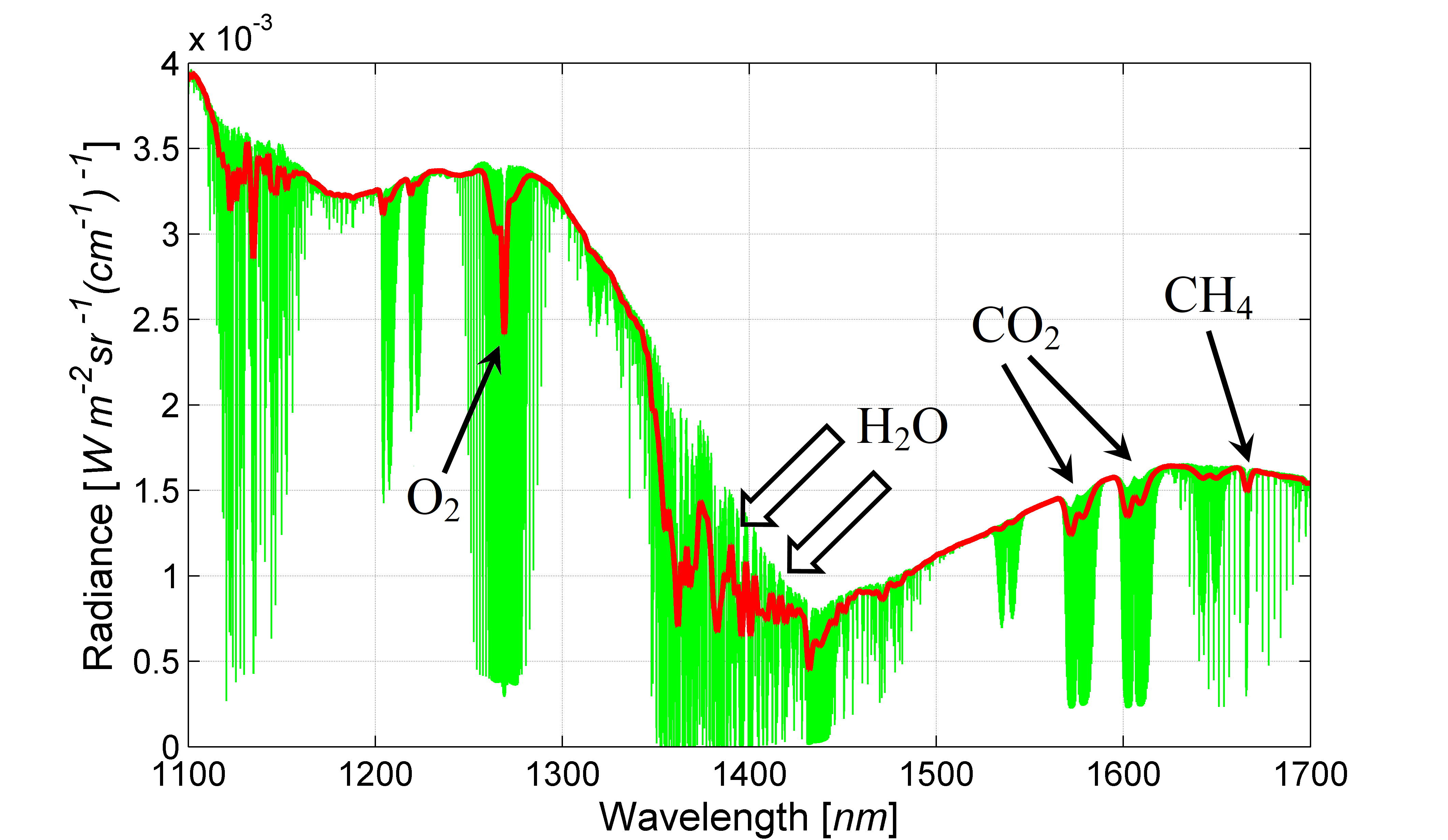}\hspace{2pc}%
\begin{minipage}[b]{26pc}
\vspace{0.3cm}
{\sffamily {\bf{Fig. 7.}} Radiance computed by using upwelling radiance proportions.}
\end{minipage}
\end{center}
\end{figure}

Figure 6 shows synthetic radiance computed by areal coverage while Fig. 7 depicts synthetic radiance computed by corresponding upwelling radiance proportions for soil, pine trees, vegetation and grass canopies. At the first glance these two figures look very similar. However, a more careful consideration reveals some discrepancies. In particular, once these two graphs from Figs 7 and 8 are overlapped, we can notice the discrepancies near $1275\,{\rm{ nm}}$, $1575\,{\rm{ nm}}$ and $1600\,{\rm{ nm}}$. This can be seen more clearly from the Fig. 8 illustrating synthetic radiance computed by areal coverage and by corresponding upwelling radiance proportions shown by magenta and blue colors.
\begin{figure}[H]
\begin{center}
\includegraphics[width=20pc]{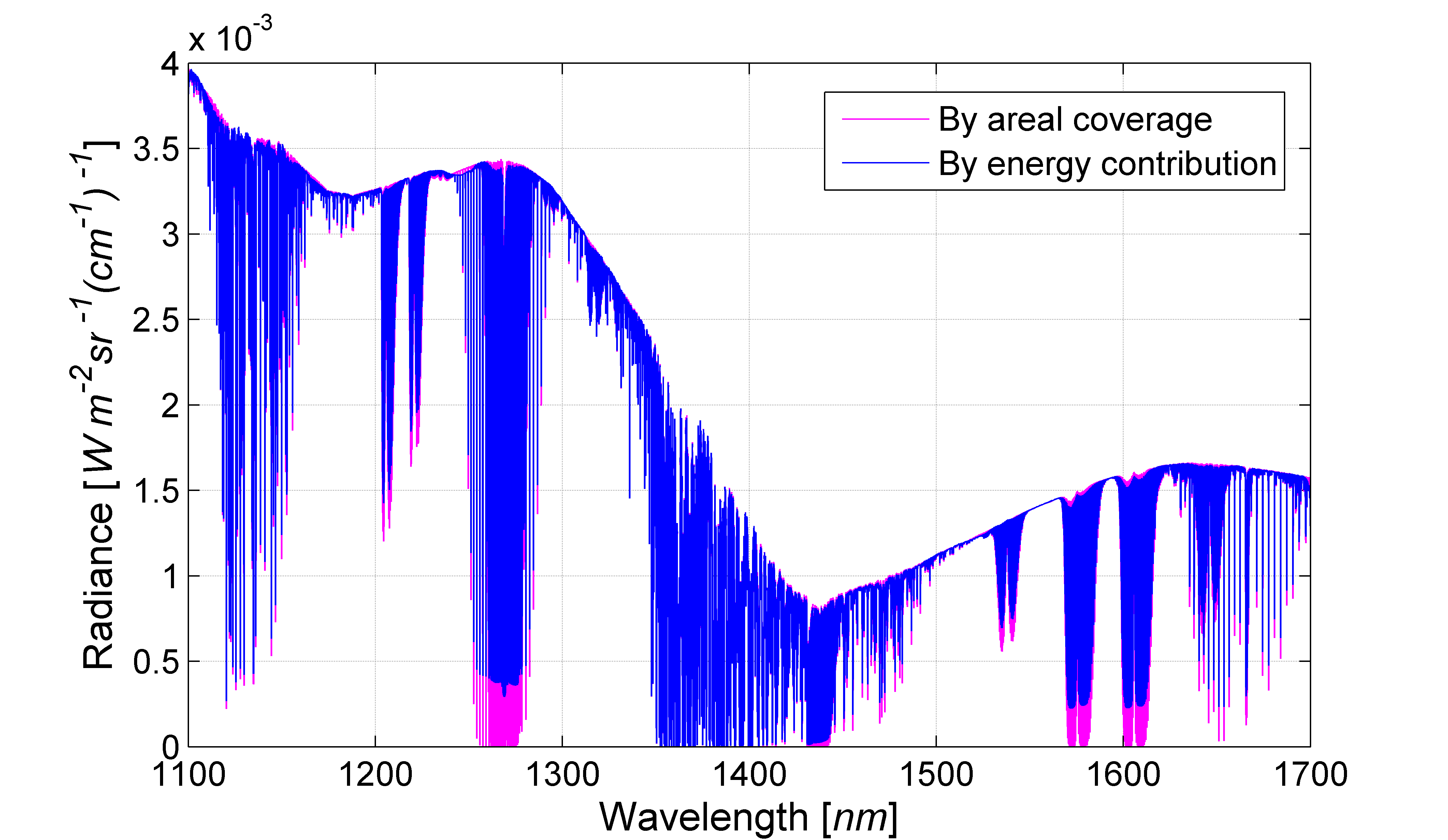}\hspace{2pc}%
\begin{minipage}[b]{26pc}
\vspace{0.3cm}
{\sffamily {\bf{Fig. 8.}} Overlapped radiance spectra obtained by both methods of computation.}
\end{minipage}
\end{center}
\end{figure}

\begin{figure}[H]
\begin{center}
\includegraphics[width=20pc]{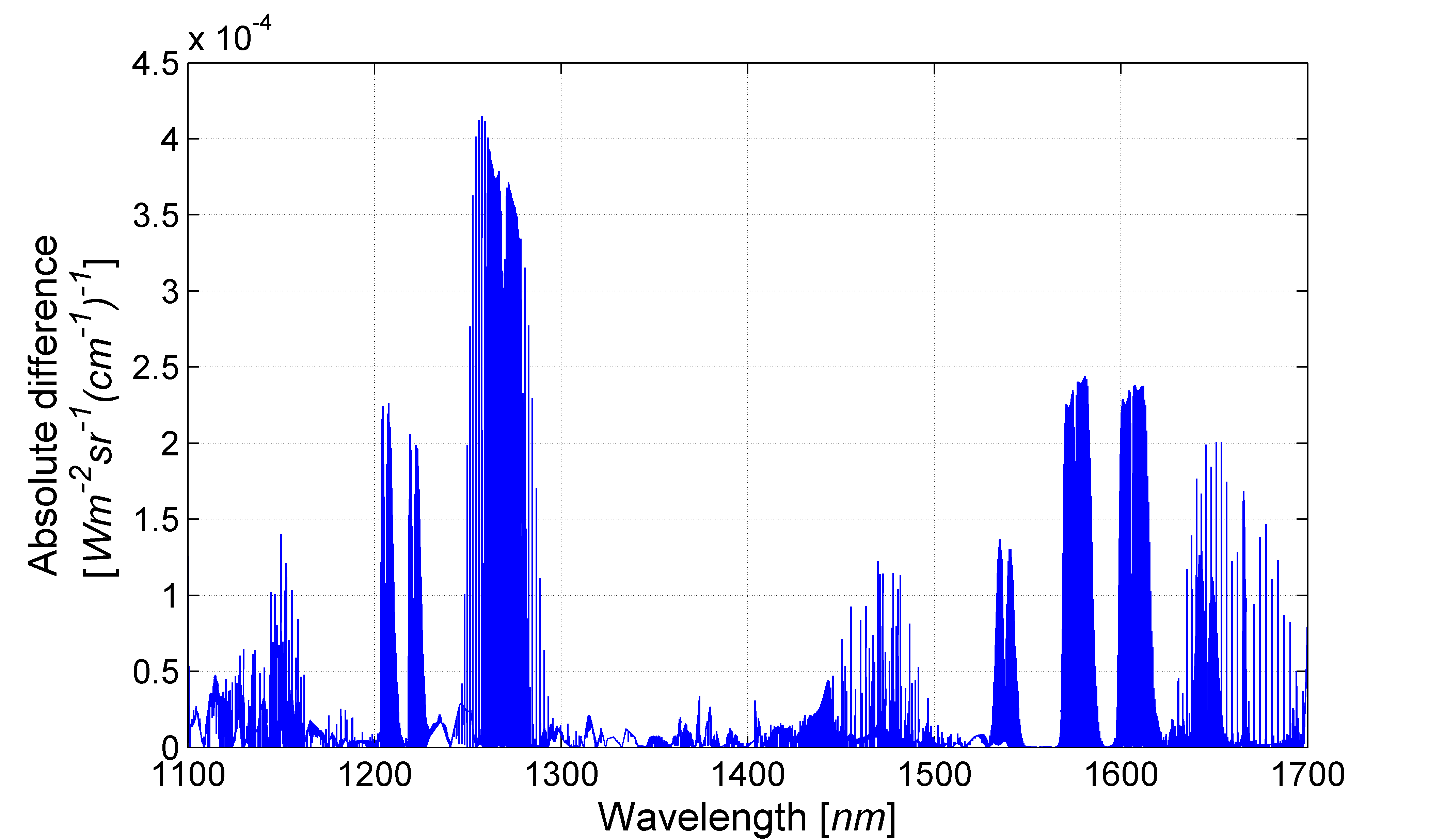}\hspace{2pc}%
\begin{minipage}[b]{24pc}
\vspace{0.3cm}
{\sffamily {\bf{Fig. 9.}} Absolute difference of radiance spectra.}
\end{minipage}
\end{center}
\end{figure}

Figure 9 shows absolute difference between these two methods. As we can see, the discrepancies are relatively large near $1275\,{\rm{ nm}}$, $1575\,{\rm{ nm}}$ and $1600\,{\rm{ nm}}$. Consequently, one may assume that these two methods are not compatible at least near these three spectral locations. However, if we take into account the instrument slit function \cite{Beirle2017, Galan1968, Roseler1966} then the results become absolutely different.
\begin{figure}[H]
\begin{center}
\includegraphics[width=20pc]{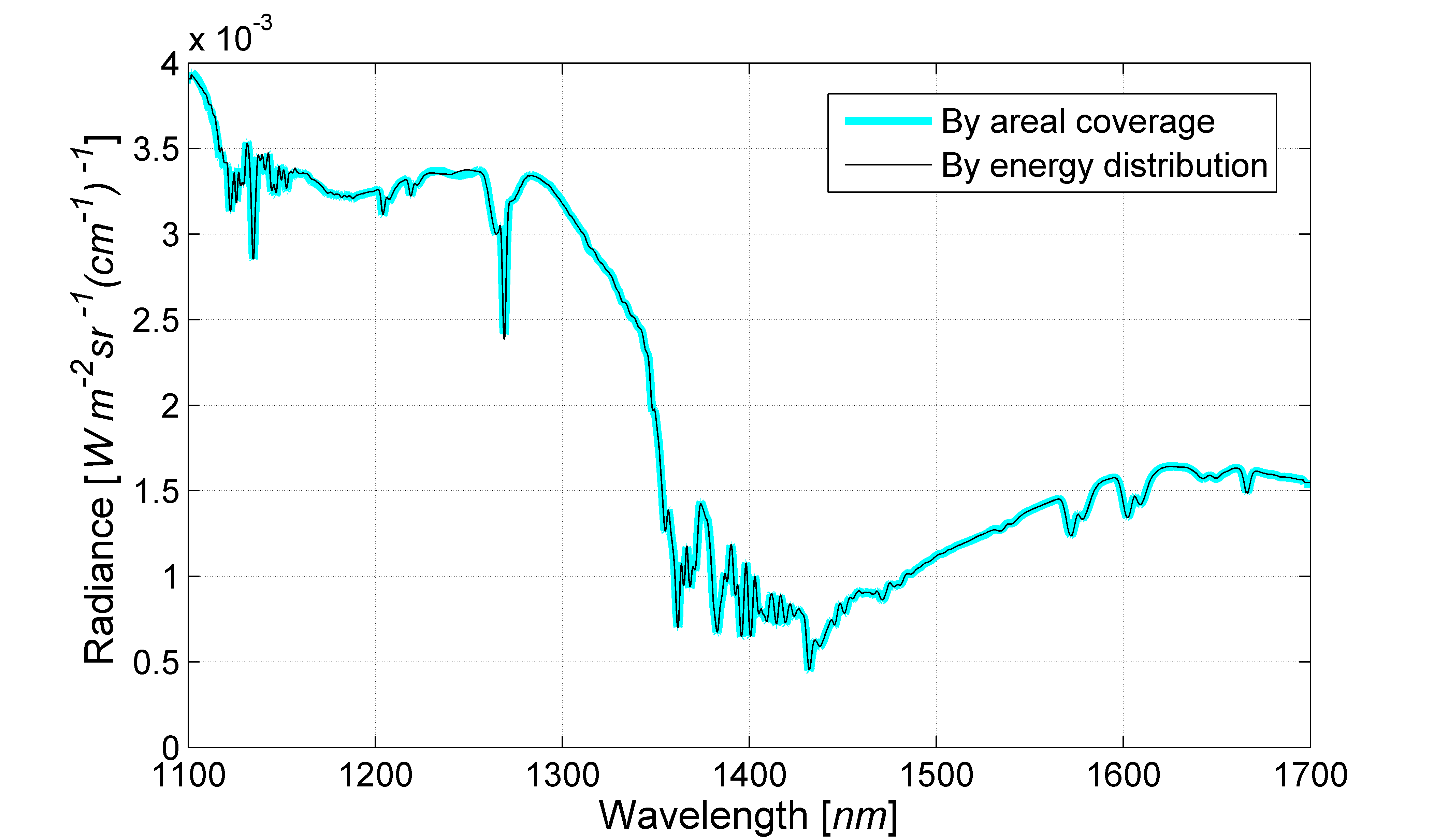}\hspace{2pc}%
\begin{minipage}[b]{24pc}
\vspace{0.3cm}
{\sffamily {\bf{Fig. 10.}} Smoothed radiance computed by both methods.}
\end{minipage}
\end{center}
\end{figure}

\begin{figure}[H]
\begin{center}
\includegraphics[width=20pc]{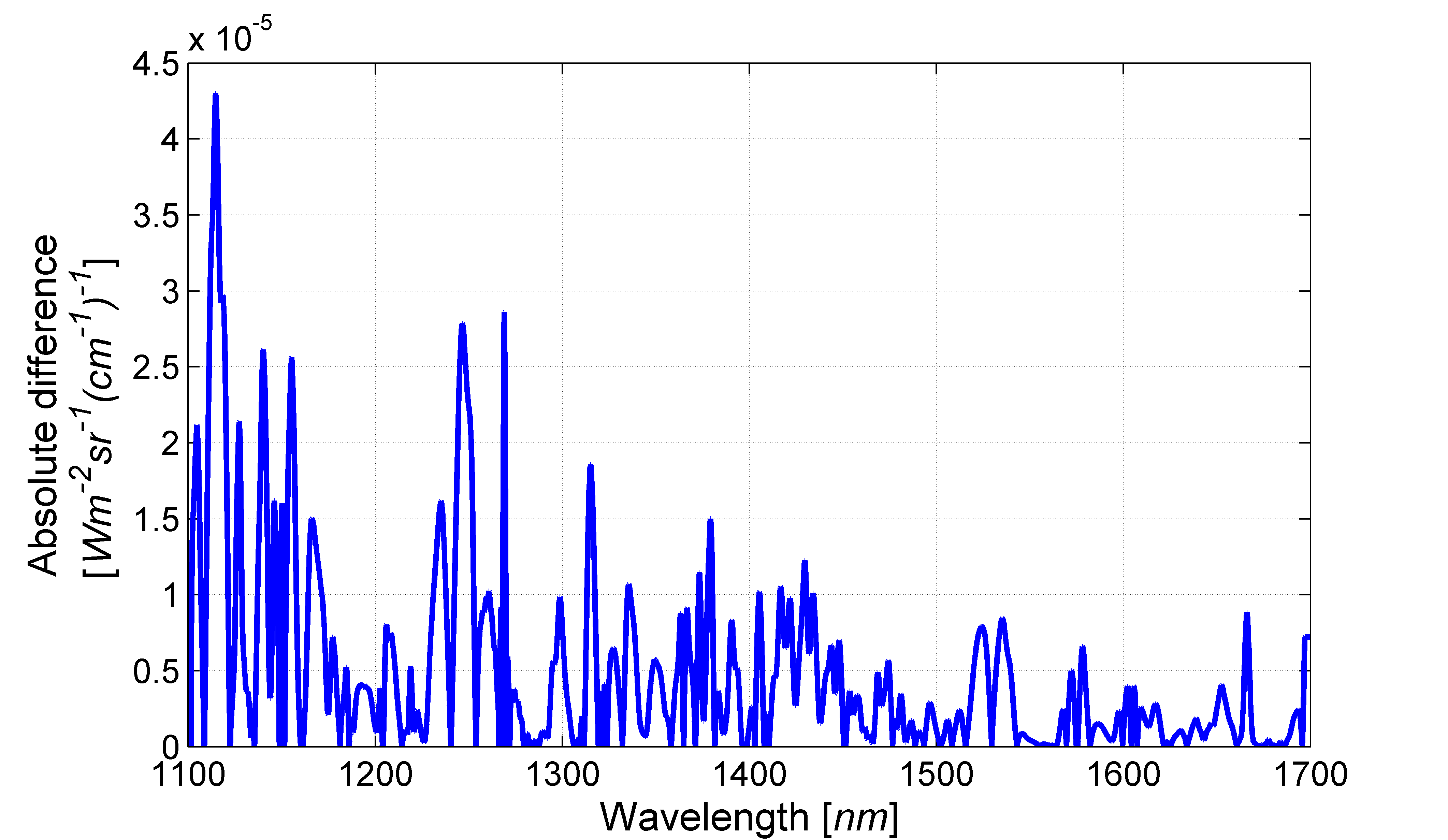}\hspace{2pc}%
\begin{minipage}[b]{26pc}
\vspace{0.3cm}
{\sffamily {\bf{Fig. 11.}} Absolute difference of radiance spectra with instrument slit function.}
\end{minipage}
\end{center}
\end{figure}

Consider Fig. 10 showing the synthetic radiance that accounts for the instrument slit function for these two methods. Visually we see no difference between curves. Figure 11 shows the absolute difference between these two curves. As we can see from this figure, the absolute difference is by more than two orders of magnitude is smaller than the actual radiance. This phenomenon can be explained from the fact that instrument slit function takes average of the highly oscillating synthetic radiance. The amplitude of oscillations may differ in these two methods. However, as an instrument slit function averages the highly oscillating synthetic radiance, both resultant curves become practically same along all wavelength range. Thus, we can see significance of the instrument slit function.

In order to verify applicability of these two methods for the entire range, we perform computation of the absolute difference as a function of contribution factor from vegetation. Figure 12a shows 3D plot of absolute difference and as we can see from this figure the worst case scenario occurs at lower frequencies and higher values of the contribution factor from vegetation. Overall, the absolute difference is by two orders of the magnitude lower (large blue area) than the smoothed upwelling radiance spectra shown in the Figure 10. However, as we can see from this Fig. 12a even the largest absolute difference is by more than $10$ times smaller. Furthermore, the red area with largest absolute difference is very narrow and negligibly small. This can be seen from Fig. 12b at the top left corner showing upper view of the same plot.
\begin{figure}[H]
\begin{center}
\includegraphics[width=28pc]{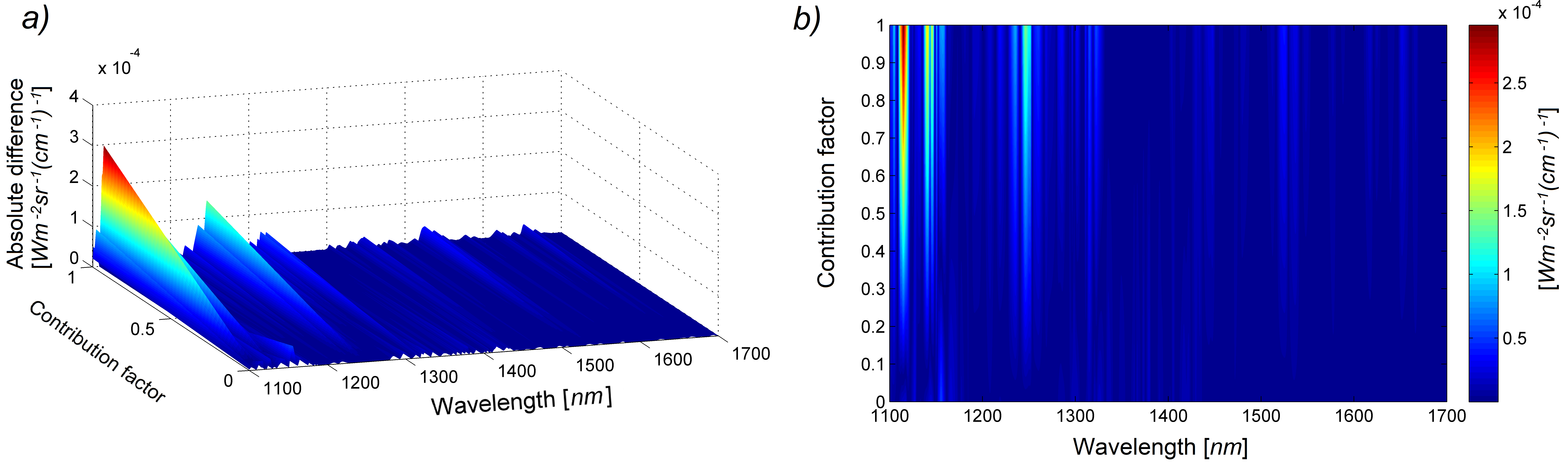}\hspace{2pc}%
\begin{minipage}[b]{26pc}
\vspace{0.3cm}
{\sffamily {\bf{Fig. 12.}} Absolute difference of radiance spectra with instrument slit function: a) side view and b) upper view.}
\end{minipage}
\end{center}
\end{figure}

Thus, the error analysis performed with the LBL radiative transfer model reveals that the method of computation based on areal coverage is completely justifiable. Both methods are equivalent for computation of the upwelling radiance. However, the method based on areal coverage is more practical. Such a good match between these two methods is due to effect of the instrument slit function that averages the highly oscillated synthetic radiance.

\section{Pixel width}

The effect of the instrument slit function showing consistency of two weighted sum methods described above should also be verified numerically more generally. In particular, it is necessary to demonstrate that the effect of the slit function remains valid regardless the instrument resolution due to a weak
dependence of the absolute error between two weighted sum methods on pixel width of the micro-spectrometer.

The size and mass limitations of a micro-spectrometer are vital in launching instrument in space because of high-cost of a payload in nano-satellite. As a result of these limitations the resolution of a micro-spectrometer are not better than $1\,nm$. Typically, spectral resolution of a modern small-size and ultralight micro-spectrometers like Argus 2000 for a space mission ranges between $1\,nm$ to $10\,nm$ \cite{Jallad2019}. A pixel width is an important parameter of a semiconductor array for detection of the light and one of the main factors that determines the spectral resolution of micro-spectrometer. Specifically, design of Argus micro-spectrometers include extended InGaAs semiconductor array consisting of more than 100 pixels in a configuration where each single pixel corresponds to the spectral counts at a particular wavenumber \cite{Jagpal2011}. Therefore, the smaller pixel width implies a higher spectral resolution of the Argus space instrument.

Figure 13 shows a set of plots of absolute errors between two weighted sum methods at pixel widths $1\,
cm^{-1}$, $5\,cm^{-1}$, $9\,cm^{-1}$, $13\,cm^{-1}$, $17\,cm^{-1}$, $21\,cm^{-1}$ and $25\,cm^{-1}$ by black, blue, red, green, brown, magenta and gray colors, respectively. The pixel width range between $1 \,cm^{-1}$ to $25\,cm^{-1}$ approximately corresponds to the spectral resolution range $0.2\,nm$ to $6\,nm$. The highest point of the black curve corresponding to the pixel width $1\,cm^-1$ is $4.99 \times 10^{-5}\,Wm^{-2}sr^{-1}(cm^{-1})^{-1}$ while the highest point of the gray curve corresponding to pixel width $25 cm^{-1}$ is $3.25 \times 10^{-5}\,Wm^{-2}sr^{-1}(cm^{-1})^{-1}$. Thus, we can see that the decrease of pixel width by factor $25$ decreases absolute difference by factor $(4.99 \times 10^{-5})/(3.25 \times 10^{-5}) \approx 1.54$ only. Therefore, our numerical results confirm a weak dependence of the absolute error on the resolution of the instrument. This signifies that two weighted sum methods retain consistency even at a relatively high spectral resolution $0.2\,nm$ of the remote sensor in detection of the light in the NIR spectral region $1100\,nm$ to $1700\,nm$.

\begin{figure}[H]
\begin{center}
\includegraphics[width=28pc]{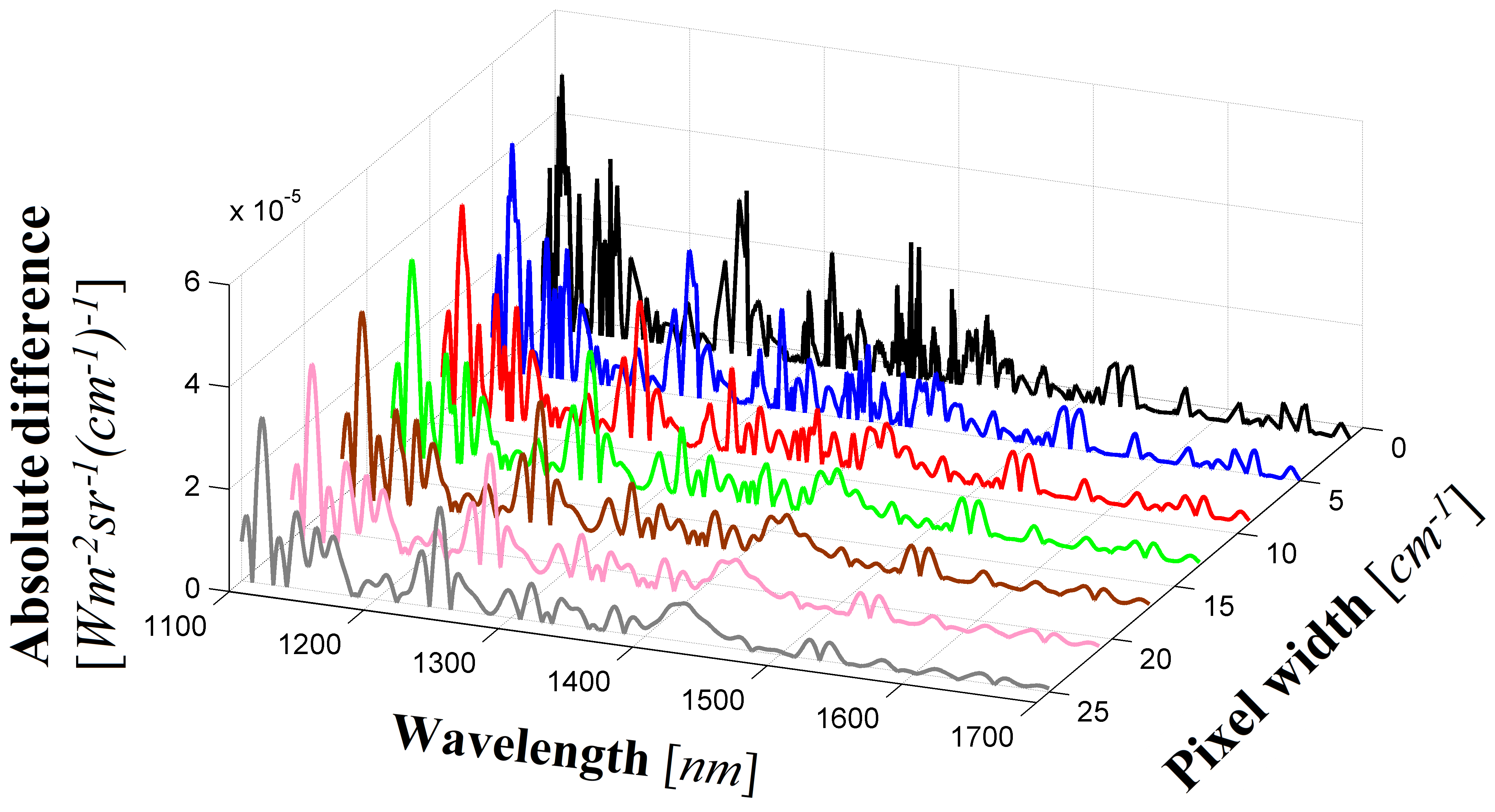}\hspace{2pc}%
\begin{minipage}[b]{26pc}
\vspace{0.3cm}
{\sffamily {\bf{Fig. 13.}} 
Absolute difference of radiance spectra at $1\,cm^{-1}$, $5\,cm^{-1}$, $9\,cm^{-1}$, $13,cm^{-1}$, $17\,cm^{-1}$, $21\,cm^{-1}$ and $25\,cm^{-1}$ pixel widths shown by black, blue, red, green, brown, magenta and gray colors, respectively.}
\end{minipage}
\end{center}
\end{figure}

\section{Conclusion}

The RE method for efficient detection of clouds and forest fires from the space applies LBL radiative 
transfer code for computation of the upwelling radiance that accounts for the wavelength dependency of 
the surface Albedo \cite{Siddiqui2021}. In particular, computation of the cumulative reflectance was based on areal coverage from each canopy. This method applies less computation since the second method based on corresponding radiance proportions requires separate computation for each canopy. Although the method of computation of the upwelling radiance based on areal coverage is more convenient in implementation, it should be validated by comparing it with method based on corresponding radiance proportions. The error analysis we performed shows that both methods are in good agreement with each other and, therefore, are virtually equivalent. In particular, our model shows that due to instrument slit function, the discrepancy between two methods near $1275\,nm$, $1575\,nm$ and $1600\,nm$ is by two orders of the magnitude smaller than the upwelling radiance. The good match between these two methods is due to instrument slit function that averages the highly oscillated upwelling radiance spectra. The proposed method can be used for retrieving space data from a micro-spectrometer operating in a NIR region. Numerical analysis we performed shows that two weighted sum methods retain consistency even at a relatively high spectral resolution $0.2\,nm$.

\section{Acknowledgments}

This study is supported by Department of Physics and Astronomy at York University, Department of Earth and Space Science and Engineering at York University, Epic College of Technology, Epic Climate Green (ECG) Inc. and Thoth Technologies Inc.

\bigskip

\end{document}